\documentclass[sigconf]{acmart} 

\copyrightyear{2025}
\acmYear{2025}
\setcopyright{cc}
\setcctype{by}
\acmConference[FAccT '25]{The 2025 ACM Conference on Fairness, Accountability, and Transparency}{June 23--26, 2025}{Athens, Greece}
\acmBooktitle{The 2025 ACM Conference on Fairness, Accountability, and Transparency (FAccT '25), June 23--26, 2025, Athens, Greece}\acmDOI{10.1145/3715275.3732015}
\acmISBN{979-8-4007-1482-5/2025/06}

\acmConference[FAccT '25]{The 2025 ACM Conference on Fairness, Accountability, and Transparency}{June 23--26,
  2025}{Athens, Greece}

\usepackage{enumitem}

\begin{document}

\title{Formalising Anti-Discrimination Law in Automated Decision Systems}

\author{Holli Sargeant}
\orcid{0000-0003-3482-7789}
\affiliation{%
  \institution{University of Cambridge}
  \city{Cambridge}
  \country{United Kingdom}}
\email{hs775@cam.ac.uk}

\author{M\aa ns Magnusson}
\orcid{0000-0002-0296-2719}
\affiliation{%
  \institution{Uppsala University}
  \city{Uppsala}
  \country{Sweden}}
\email{mans.magnusson@statistik.uu.se}

\renewcommand{\shortauthors}{Sargeant and Magnusson}

\begin{abstract}
Algorithmic discrimination is a critical concern as machine learning models are used in high-stakes decision-making in legally protected contexts. 
Although substantial research on algorithmic bias and discrimination has led to the development of fairness metrics, several critical legal issues remain unaddressed in practice. 
The paper addresses three key shortcomings in prevailing ML fairness paradigms: (1) the narrow reliance on prediction or outcome disparity as evidence for discrimination, (2) the lack of nuanced evaluation of estimation error and assumptions that the true causal structure and data-generating process are known, and (3) the overwhelming dominance of US-based analyses which has inadvertently fostered some misconceptions regarding lawful modelling practices in other jurisdictions.
To address these gaps, we introduce a novel decision-theoretic framework grounded in anti-discrimination law of the United Kingdom, which has global influence and aligns closely with European and Commonwealth legal systems. 
We propose the ``conditional estimation parity'' metric, which accounts for estimation error and the underlying data-generating process, aligning with UK legal standards. 
We apply our formalism to a real-world algorithmic discrimination case, demonstrating how technical and legal reasoning can be aligned to detect and mitigate unlawful discrimination. 
Our contributions offer actionable, legally-grounded guidance for ML practitioners, policymakers, and legal scholars seeking to develop non-discriminatory automated decision systems that are legally robust.
\end{abstract}

%% The code below is generated by the tool at http://dl.acm.org/ccs.cfm.



\maketitle

\section{Introduction}

Instances of large-scale failures of supervised machine learning (\textbf{SML}) based decision systems, from disproportionately harming vulnerable people in algorithmic immigration assessments~\cite{forster2022,oxkul2023}, unfair welfare eligibility assessments~\cite{eubanks2018,senecat2023,Lighthouse2024}, to perpetuating racial biases within criminal justice systems~\cite{zilka2022,adams2024}, have spurred a rich literature on algorithmic bias and discrimination. 
Despite such literature from the ML and legal communities focusing on fairness metrics intended to identify and address algorithmic bias~\cite{mayson2019,ho2020,xiang2021,barocas2016,hacker2018,wachter20212,xenidis2020,hellman20201,gillis2022,adams2022,weerts2023}, several critical legal issues remain unaddressed in practice. 
Practitioners, especially model developers and decision-makers, continue to lack precise guidance on effectively avoiding and mitigating unlawful algorithmic discrimination in compliance with anti-discrimination laws across various jurisdictions.
This paper explicitly identifies and prioritises three primary shortcomings prevalent in current approaches that, if ignored, undermine both the lawfulness and practical effectiveness of fairness interventions.

\textit{First}, while algorithmic fairness research largely emphasises discrimination as statistical disparities in predicted outcomes for binary, marginalised groups.
The common conflation of fairness and legal non-discrimination has led to a predominant focus on outcome disparity as the main indicator of unfairness, yet \emph{unlawful} discrimination is both broader and more detailed. 
This gap creates a critical mismatch between what ML methods identify as unfair and what anti-discrimination law recognises as unlawful, rendering many fairness interventions inadequate, counterproductive, or even~unlawful.

\textit{Second}, a persistent disconnect exists between technical modelling practices and the requirements of anti-discrimination law, resulting in legal blind spots and methodological oversights. 
Conventional technical approaches frequently overlook important aspects of statistical uncertainty and estimation error, fail to account for the underlying data-generating process (\textbf{DGP}), or assume that the true causal structure behind discrimination is already known. 
In reality, the outputs of SML models are approximations, subject to inherit sampling variability, label noise, data limitations, and model misspecification. 
These methodological issues are not academic abstractions -- they carry concrete legal and practical consequences.
%Yet these have significant legal and practical consequences. 

\textit{Third}, the predominance of analysis of fairness and discrimination in ML from the United States (\textbf{US}), lack of non-US ML examples~\cite{quy2022}, and limited legal scholarship translating these concepts across other jurisdictions, has inadvertently fostered a series of misconceptions on what would be \emph{unlawful} algorithmic discrimination.
Few papers have engaged with anti-discrimination laws outside the US~\cite[see e.g.,][]{xenidis2020,lattimore2020,wachter20211,weerts2023,kiesow2023,langenbucher2023}, and fewer still in the United Kingdom (\textbf{UK})~\cite[see][]{sanchez2020,adams2022,kelly2021}.
By avoiding the nuanced legal realities of other jurisdictions, models designed to comply with US laws may breach UK laws or those in comparable jurisdictions.
English common law is either in force or is the dominant influence in 80 legal systems that govern approximately 2.8 billion people, not including the US~\cite{CIA}.
In particular, UK anti-discrimination law is very similar to numerous Commonwealth and common law jurisdictions, including Australia~\cite{AHRC2014}, Canada~\cite{CanadaHRA}, India~\cite{IndiaConst}, New Zealand~\cite{NZHRA}, South Africa~\cite{SADL}, and the pending bill in Bangladesh~\cite{BADB2022}.
European Union (\textbf{EU}) law also has broadly the same discrimination law as it evolved in parallel during the UK's membership~\cite{EUCFR}.

As new AI regulations emerge worldwide to prevent discriminatory practices, the need for a more jurisdictionally nuanced understanding of unlawful discrimination decision-making based on SML is imperative~\cite{EUAIAct,EO14110}.
Our paper addresses this gap by providing a rigorous analysis of UK anti-discrimination law, correcting certain mischaracterisations, and establishing a more accurate foundation for developing non-discriminatory ML in the UK and related jurisdictions.

\paragraph{Contributions} 
To address these gaps, our paper makes four core contributions at the intersection of automated decision-making, fairness, and anti-discrimination doctrine.
\begin{enumerate}
    \item We introduce a formalisation of UK anti-discrimination doctrine within a decision-theoretic framework to provide a legally informed approach to SML for automated decisions.
    \item We introduce the concept of the true data-generating process (\textbf{DGP}) as a theoretical construct, allowing for a systematic evaluation of the legitimacy of prediction targets \(y\) and features \(x\) in SML models.
    \item We propose ``conditional estimation parity'' as a new, legally informed target to minimise the legal and practical effects of estimation error in SML models.
    \item We provide recommendations on creating SML models that minimise the risk of unlawful discrimination in automated decision-making in the UK and related jurisdictions.
\end{enumerate}

The paper is structured as follows:
Section~\ref{sec:background} outlines our notation and formalisations of automated decision-making, surveys the algorithmic fairness literature, and introduces our functional analysis of UK anti-discrimination law and decision-theoretic approach to SML.
Section~\ref{sec:prohibited} examines what kinds of discrimination the law prohibits, clarifying the distinctions between algorithmic direct and indirect discrimination. 
Section~\ref{sec:allowed} analyses the instances where the law allows certain types of differential treatment, introducing the concepts of legitimacy, the true DGP, and estimation parity. 
In Section~\ref{sec:liable}, we discuss how decision-makers can be found liable for unlawful discrimination, linking statistical disparities and legal causation to show how \emph{prima facie} discrimination can emerge from model outcomes. 
Section~\ref{sec:defences} turns to possible justifications for certain disparate outcomes.
To connect our theoretical framework to judicial reasoning, in Section~\ref{sec_case_study} we apply our approach to a real-world algorithmic discrimination case.
Section~\ref{sec:conclusion} concludes. 

\section{Automated Decisions and Discrimination} \label{sec:background}
\subsection{Automated Decision-Making}

Let \(x_{i}\in \mathbb{R}^p\) be a vector of observed attributes for individual \(i\). 
A decision-maker must choose a decision \(a\in\mathcal{A}\), where \(\mathcal{A}\) is closed. 
We assume the decision-maker wants to decide based on a future outcome \(y_i \in \mathcal{Y}\) for individual \(i\). Further, we assume \(\mathcal{Y}=\mathbb{N}\), which can be relaxed. 
Decision-making under uncertainty has long been studied in statistical decision theory~\cite{savage1956,degroot1970optimal,berger1985statistical,parmigiani2010}.
In a decision setting, let \(u(y,a)\) be a utility function that summarises the \emph{utility} for the decision-maker, where the optimal decision under uncertainty is 

\begin{equation}
a^\star = \underset{a\in\mathcal{A}}{\arg\max}  \sum_{y\in\mathcal{Y}} u(y,a) p(y|a,x)\,.
\label{eq_optimal_action}    
\end{equation}

The decision-maker usually neither knows \(y_i\) nor \(p(y|a,x)\) at the time of the decision. 
Hence, the decision must be based on \(x_i\). Here, we drop $a$ for simplicity, assuming that $a$ does not affect $y$. 
In an SML setting, a prediction model \(\hat{p}(y|x)\) is trained to compute the predicted probability distribution (pmf) \(\hat{\boldsymbol{\pi}}_i = \hat{p}(y|x_i)\) for individual \(i\), with the support on \(\mathcal{Y}\). 
Further, let \(\hat{y}(\hat{\boldsymbol{\pi}}_i)\in\mathcal{Y}\) be the classification made based on \(\hat{\boldsymbol{\pi}}_i\). 
In simple settings, the decision can be formulated as a decision function \(d(\hat{\boldsymbol{\pi}}_i) \in \mathcal{A}\) that is used to choose an appropriate action based on \(\hat{\boldsymbol{\pi}}_i\). 
In the binary \(y\) and \(a\) case, it reduces to a simple threshold \(\tau\), i.e., \(d(\hat{\pi}) = I(\hat{\pi}\leq\tau)\),
where \(I\) is the indicator function and \(\hat{\pi}_i = \hat{p}(y=1 \mid x_i)\). 
We often train a model \(\hat{p}(y|x)\) based on previous data \(D=(\mathbf{y},\mathbf{X})\), drawn from a population \(p(y,x)\), where both \(x_i\) and \(y_i\) are known. 
Replacing \(p(y|x_i)\) with the predictive model \(\hat{p}(y|x_i)\) in Eq.~\ref{eq_optimal_action} gives an optimal decision using the prediction model.

\subsection{Algorithmic Fairness}

Algorithmic fairness metrics generally measure prediction disparities across groups with different legally protected characteristics commonly identified in datasets, including gender and race~\cite{hardt2016,corbett2017,narayanan2018,verma2018,mulligan2019,binns2021,mehrabi2021,mitchell2021}.
This research has resulted in several proposals, including statistical metrics to assess the fairness of individual predictive models~\cite{verma2018,speicher2018,carey2023,caton2023}, fairness for model auditing~\cite{kim2017,raji2020,kasy2021,mokander2023,oneil2023}, and fairness constraints on models~\cite{corbett2017,zafar2019,Heidari2019,Xu2020,becker2023}. 
We outline two core metrics that are relevant to our work -- statistical parity and conditional statistical parity.
To define these metrics in our notation, we separate \(x_i\) into protected and legitimate features \(x_i=(x_{pi},x_{li})\); we drop \(i\) to simplify notation.
Here, \(x_p \in \mathcal{C}\) indicates protected attributes, with \(\mathcal{C}\) being the set of different categories or groups. 

\textbf{Statistical parity}, or demographic parity, is one of the central algorithmic fairness metrics~\cite{corbett2017,verma2018,narayanan2018,mehrabi2021}.
For statistical parity to hold, it requires that
\begin{equation}
\mathbb{E}_x\left[\hat{p}(y|x) \mid x_p\right] = \mathbb{E}_x\left[\hat{p}(y|x)\right]\,, 
\label{eq_stat_parity}
\end{equation}
such that the model predictions, in expectation over \(x\), need to be the same for the different groups~\cite{corbett2017,verma2018}. 
Given that the decision function \(d(\boldsymbol{\pi})\) is the same for the different groups, statistical parity results in equal decisions across~those~groups. 

\textbf{Conditional statistical parity} extends statistical parity to account for legitimate features \(x_l\). 
It requires that
\begin{equation}
\mathbb{E}_x\left[\hat{p}(y|x) \mid x_l, x_p\right] = \mathbb{E}_x\left[\hat{p}(y|x)\mid x_l\right]\,, 
\label{eq_cond_stat_parity}
\end{equation}
so there should be no difference in model predictions or decisions between groups given by the protected attribute, conditional on legitimate features \(x_l\)~\cite{corbett2017,verma2018,castelnovo2022}. 

Other related group comparison metrics include error parity, balanced classification rates, and equalised odds~\cite{hardt2016,corbett2017,verma2018,narayanan2018,mehrabi2021,corbett2023}.
Individual approaches to parity have also considered whether otherwise identical individuals are treated differently if they have different protected attributes~\cite{dwork2012,kilbertus2018,binns2020}.
Concepts from causal inference and counterfactual reasoning have been proposed to measure outcome consistency for individuals across protected groups~\cite{kusner2017,kilbertus2017,russell2017,zhang2018fairness,chiappa2019,wu2019,nilforoshan2022causal,alvarez2023counterfactual}.
It is not in the scope or aim of this paper to evaluate these numerous algorithmic fairness metrics.

\subsection{Anti-Discrimination Law and Decision-Theoretic Approach}

Identifying a discriminatory AI system has occurred in a variety of contexts, often without identifying whether it is \textit{unlawfully} discriminatory.  
A non-lawyer may be surprised by several types of unfair behaviours that are not legally prohibited and some inconspicuous decisions that result in unlawful discrimination~\cite{khaitan2015}.  
As a term that has entered the common vernacular, despite its specific meanings in several contexts, it has led to the conflation of discrimination and unlawful discrimination.
Importantly, anti-discrimination law only applies to a select group of duty-bearers~\cite{khaitan2015}.
For example, an individual choosing not to be friends with people based on their sexuality or choosing not to marry someone based on their race is not unlawful~\cite{emens2009,khaitan2015}.
Despite certain actions appearing unreasonable or unfair, they are not always prohibited. 
Consider an algorithm that rejects a loan application because the applicant uses an Android phone rather than an iOS device~\cite{aggarwal2021,langenbucher2023}.
While it may show a correlation to the applicant's income, one may agree that it is unfair because it does not reflect the individual likelihood of defaulting on a loan and instead penalises the individual for how much money they spent on a mobile device.
However, under UK law, it would not be \textit{unlawful} because spending habits, income, or even poverty, are not protected attributes~\cite{poverty2022}.
Therefore, to engage in a functional analysis of UK anti-discrimination laws in automated decision systems, it is helpful to frame its legal elements. 

%\subsubsection{Functional Analysis of Anti-Discrimination Law}

Anti-discrimination law operates through two main functions: an \emph{ex ante} role and an \emph{ex post} role.
First, it sets rules that define prohibited conduct and the contexts in which these rules apply, offering guidance on acceptable, prohibited, or potentially justifiable actions (rule articulation). 
The elements that define the \textit{rule} against unlawful discrimination are (1) protected contexts, (2) protected characteristics, and (3) prohibited conduct. 
Second, if a rule is violated, the law takes on a different function and sets conditions for discrimination liability (liability).
If an action appears to be prohibited by the rule articulation function, the legal analysis shifts to whether there is \textit{liability} and if it can be excused. 

In most algorithmic fairness literature, the primary, or often the sole, consideration is whether the ML outputs result in disparate predictions or decisions for a protected group~\cite{hardt2016,verma2018,narayanan2018}. 
Such approach only considers aspects of the rule articulation function of anti-discrimination law.
However, we take a wholistic, legally-informed approach:
\begin{enumerate}
    \item What types of discrimination does the law prohibit? 
    \item What types of discrimination does the law allow? 
    \item How can a decision-maker be liable for unlawful discrimination?
    \item When can a decision-maker be excused for unlawful discrimination?
\end{enumerate}

We formalise UK anti-discrimination law in a decision-theoretic framework that provides a systematic approach for the decision-maker to make optimal ML design choices under uncertainty and under legal constraints. 
Our decision-theoretic view, where actions map states to outcomes, each with associated utilities, enables a formal assessment of modelling choices under UK anti-discrimination~law. 

\section{What types of discrimination does the law prohibit?} \label{sec:prohibited}

Unlawful and prohibited discrimination is defined by (1) protected contexts, (2) protected characteristics, and (3) prohibited conduct.
Protected contexts are defined by the imposed duties on government, employers, landlords, providers of goods and services~\cite{EqualityAct}. 
The protected characteristics under the Equality Act are age, disability, gender reassignment, marriage and civil partnership, pregnancy and maternity, race, religion or belief, sex and sexual orientation~\cite[s 4]{EqualityAct}.
Two types of conduct are prohibited: \textit{direct} discrimination and \textit{indirect} discrimination.
%Direct discrimination occurs when an individual is treated less favourably than another based on a protected characteristic~\cite[s 13]{EqualityAct}. 
%Indirect discrimination refers to a policy, criterion, or practice (\textbf{PCP}) that disproportionately disadvantages a group with a particular protected attribute compared to those without~\cite[s 19]{EqualityAct}. 

\subsection{Algorithmic Direct Discrimination}

Less favourable treatment of an individual in a protected context based on one or multiple protected characteristics is unlawful direct discrimination~\cite[s 13(1), 14]{EqualityAct}. 
Where a model \(\hat{p}(y|x)\) uses a protected attribute \(x_p\), and there is a difference in predictions between the protected groups defined by \(x_p\) that results in less favourable treatment for an individual in that group, this risk arises. 

In US literature, the prevailing view is that the analogous type of direct discrimination, ``disparate treatment'', will be challenging to prove in an algorithmic context~\cite{barocas2016}.
However, the dominance of US legal framing has created an assumption that direct discrimination is similarly ``likely to be less important''~\cite{wachter20212},\cite[see also][]{kelly2021,xenidis2020,Zuiderveen2020}.
This assumption is incorrect.
As \citet{adams2022} explain, ``the scope of direct discrimination is significantly wider than that of disparate treatment.''

First, UK direct discrimination focuses on whether a protected characteristic is the reason for less favourable treatment~\cite{Essop}; it does not take a formalistic view of whether \(x_p\) is considered in the decision-making in the ``input-focused disparate treatment'' doctrine of the US~\cite[cf.][]{gillis2022}.
The formalistic approach in the US resulted in computer science literature incorrectly encouraging the removal of protected attributes when designing ML~\cite{romei2014,johndrow2019,grabowicz2022},\cite[cf.][]{Yang2020,hellman20201}.
However, in the UK there is no general prohibition on the knowledge or consideration of a protected characteristic, and even if a model ignores \(x_p\), in practice, direct discrimination can also be based on a criterion that is some ``indissociable'' proxy which has an ``exact correspondence'' to the protected characteristic \(x_p\)~\cite{court2017,court2018}.
Formally, we can define an exact proxy as a feature \(\tilde{x}_{p}\) with an exact correspondence with~\(x_p\). 
UK Courts have held exact proxies to include the criterion of statutory retirement age that differed between men and women as a proxy for sex~\cite{lords1990}, the criterion of marriage was historically indissociable from heterosexual orientation~\cite{Preddy}, or pregnancy as an exact proxy to the female sex~\cite{lords1995,tribunal1996}.
This approach to direct proxy discrimination also applies under EU law~\cite{weerts2023,Dekker,Hay, Maruko}.
If a model uses such features, it would have the effect of using an exact proxy \(\tilde{x}_{p}\) that could be the basis for a direct discrimination claim.
Therefore, removing \(x_p\) will not avoid liability for unlawful direct discrimination.

Further, from a technical perspective simply removing protected characteristics may reduce accuracy and utility~\cite{Zhang2017,khani2021}, and does not remove the risk of bias or disparity~\cite{dwork2012,lipton2018,kleinberg2019}.
If the inclusion of~\(x_p\)~or~\(\tilde{x}_{p}\)~improves the model accuracy without resulting in less favourable treatment for protected individuals, it may avoid direct discrimination.
There is an absence of any legal guidance in the UK on the use of protected attributes in automated decision-making.\footnote{Although in the context of data processing, it may be lawful to use data on personal characteristics if necessary to identify or review equality of opportunity or treatment between groups of people, or to prevent or detect other unlawful acts~\cite[Sch 1]{DPA}.} 
Pending further legal guidance, it is important to carefully consider the effects of including~\(x_p\)~or~\(\tilde{x}_{p}\).

Second, intention is irrelevant to direct discrimination, ``no hostile or malicious motive is required''~\cite{Essop,lords1990,EandJFS}. 
Importantly, this diverges from US law and highlights that intention is immaterial to UK direct discrimination and to consideration of intentional proxy discrimination~\cite[cf.][]{prince2020,tschantz2022}.

\subsection{Algorithmic Indirect Discrimination}
Where a facially neutral provision, criterion, or practice (\textbf{PCP})  disproportionately disadvantages individuals with a protected attribute, it is unlawful indirect discrimination~\cite[s 19(1)]{EqualityAct}.
The UK Supreme Court (\textbf{UKSC}) explains that it ``aims to achieve equality of results in the absence of such justification''~\cite[para 25]{Essop}. 
Such PCP is discriminatory if it applies to persons with the protected characteristic and puts, or would put, persons with that characteristic at a particular disadvantage when compared to those without such attributes~\cite[s 19(2)]{EqualityAct}.
There is no requirement that the PCP puts every member of the protected group at a disadvantage, nor the need to prove the reason for the disadvantage~\cite{Essop}.
Further, even individuals without the relevant protected characteristic who suffer from the same disadvantage of a discriminatory PCP as those with the protected characteristic can also bring a claim for indirect discrimination (also known as ``discrimination by association'')~\cite[established in EU case law][]{CHEZ},\cite[adopted in the UK in 2024,][s 19A]{EqualityAct}.
Therefore, the definition of indirect discrimination is broader under UK law than in the US and, as will be discussed below, the thresholds for proving liability is lower in the UK than under US doctrine.  
The scope of indirect discrimination means there are many potential avenues for it to arise in an algorithmic context.

\section{What types of discrimination does the law allow?} 
\label{sec:allowed}

\subsection{Legitimacy of True Differences} \label{sec:legitimacy_of_true_differences}

Anti-discrimination law in the UK and related jurisdictions aims to mitigate or eliminate any material disadvantages between protected groups and their comparators.
It recognises the legitimacy of true and relative differences between individuals.
Sometimes it overrides legitimate differentiation between people, and other times it expressly allows differentiation. 
Such nuance is often ignored in algorithmic fairness literature with two primary consequences: (1) the focus on parity has become disconnected from legal realities by failing to analyse individual differences, and (2) the failure to distinguish ground truth and estimations in SML-based decision-making avoids crucial legal nuance and can lead to unlawful discrimination.

Direct discrimination requires formal equality of treatment.
It specifically prohibits less favourable treatment of a person \textit{based on} a protected characteristic, even if there is a true difference in their risk, ability, or merit.
In addition to prohibiting prejudice, direct discrimination acknowledges that there are certain protected characteristics that should not be used as the basis for less favourable treatment even though that characteristic may be relevant.
Although gender, pregnancy, certain disabilities, among other things may be relevant to a person's job performance, the law prohibits discrimination on those grounds~\cite{khaitan2015}.
For example, until 2012 insurers could use sex as a determining factor in risk assessments based on actuarial and statistical data.
However, the European Court of Justice overturned this rule~\cite{TestAchats}, arguing that the use of sex as a risk factor should not result in differences in individuals’ insurance premiums and benefits~\cite{EA2012}. 
Instead of using sex as a variable, insurers need to consider information that is not protected and closer to predicting the true question of risk. 
Direct discrimination ``allows only carefully defined distinctions and otherwise expects symmetry''~\cite{EandJFS}.
These defined distinctions are important. 
For instance, insurance decisions that might otherwise be construed as discriminatory -- specifically concerning gender reassignment, marriage, civil partnership, pregnancy, and sex discrimination -- are permissible if they are based on reliable actuarial data and executed reasonably~\cite[Sch 9 s 20]{EqualityAct}. 
Similarly, financial services can ``use age as a criterion for pricing risk, as it is a key risk factor associated with for example, medical conditions, ability to drive, likelihood of making an insurance claim and the ability to repay a loan''~\cite[para 7.6]{agediscrimnotes}.
Similar statutory exemptions are found in anti-discrimination laws in the EU~\cite[art 2]{EU2002}, Australia~\cite[s 30-47]{AUSDA}, Canada~\cite[s 15]{CanadaHRA}, New Zealand~\cite[s 24-60]{NZHRA} and South Africa~\cite[s 14]{SADL}.
These exemptions legally recognise that certain group distinctions, particularly those involving risk assessment, are relevant and necessary for the equitable operation of such services.

On the other hand, indirect discrimination recognises that formal neutrality or even the application of formal equality can create disadvantages to protected groups. 
Avoiding indirect discrimination may sometimes require providing opportunities in a manner that only discriminates when accounting for legitimate individual and group differences -- aiming for substantive equality~\cite{vos2020,xenidis2020,wachter20211}.
Consider a scenario where a store manager imposes a minimum height requirement for all employees, justifying the policy by the need for all employees to equally be able to access supplies stored on high shelves. 
Although the rule may appear neutral, it would, in practice, disproportionately disadvantage women, whose average heights are lower than those of men. 
Indirect discrimination recognises the true height differences between certain groups. 

%While beyond the scope of this analysis there are several relevant moral evaluations of legitimate differences, for instance, that just distributions should be according to some merit~\cite{aristotle2000,gosepath2015}, or inequalities may be permissible to the extent it is of the greatest benefit to the least advantaged in society~\parencites[][]{rawls1971}[][]{rawls2001}[in algorithmic fairness context see][]{franke2024}[][]{binns2024}.

\subsection{Ground Truth, True Data Generating Process and Estimation}\label{sec_true_data}

Similar to related work on measurement bias and construct validity~\cite{jacobs2021}, we explore a fundamental and overlooked limitation of legal frameworks in algorithmic discrimination: law is predicated on an individualistic model that often requires a threshold of identifiable harm before legal action can be triggered. 
It is not inherently designed to work with approximations or mitigating systemic level harms. 
We outline a critical intervention that emerges from comparing true underlying risks and  estimation of that risk. 
Therefore, an important aspect from the legal perspective that is overlooked in the algorithmic fairness literature, but is a standard theoretical framework  in statistics, the distinction between a true DGP and the estimated model \(\hat{p}(y|x)\).
To formalise, we assume that there exists a true DGP, \(D \sim p(y,x)\), where \(D_i=(y_i,x_i)\). 
Further, we use \(p(y|x^{\text{true}}_i)\) to denote the true probability (pmf) for individual \(i\), given the true features~\(x^{\text{true}}_i\). 

We make multiple observations on the role of the \textit{true} model and its use in connecting predictive modelling and legal reasoning.

First, understanding the limits of predictive models is crucial to explore inherent uncertainties and limitations in predictions. 
The true model is, in practice, never observed or known. 
When developing \(\hat{p}(y|x)\), the target is often to select the model with the best predictive performance, which is closely connected to the true DGP~\cite{shao1993linear,bernardo1994bayesian,vehtari2002bayesian,vehtari2012survey,vehtari2017practical}.
The true model may include features in \(x^{\text{true}}_i\) that are not observed in the data, sometimes referred to as an \(\mathcal{M}\)-open setting, i.e., when the true model is not included in the set of candidate models~\cite{bernardo1994bayesian,vehtari2012survey}.

Second, we assume that \(p(y|x_i)\) is a probability distribution over \(\mathcal{Y}\), introducing some level of aleatoric uncertainty in the true underlying process~\cite{o_hagan2004,Hullermeier2021,tahir2023}.
This means that perfect prediction of \(y_i\) may not be possible, even with knowledge of the true DGP. 
The distinction between aleatoric and epistemic uncertainty is important from a legal perspective. 
The reason is simple: the uncertainty coming from estimation and modelling is the (legal) responsibility of the modeller, while the aleatoric uncertainty can instead be considered a true underlying general risk. 

Third, the true DGP connects to judicial legal reasoning. 
Courts must engage theoretically with legal and normative conceptions of what constitutes unlawful discrimination and what is justifiable. 
Judges consider legitimacy, proportionality, and necessity when evaluating actions, and hypothetical alternatives, that led to discriminatory treatment or outcomes.
Although case law does not always represent ground truth, courts can operate as an oracle. 
It may not pinpoint what the perfect decision should have been, but courts will engage in a similar theoretical process of reasoning about the decision-making process to the true DGP to understand whether the actions were justified or unlawful.
We explain legal reasoning within this framework throughout the paper and in a case on unlawful discrimination in algorithmic decision-making~(Section~\ref{sec_case_study}).

\subsection{Estimation Parity} \label{sec_estimation_parity}

Therefore, it is legally important to distinguish between a true difference and an estimated one. 
We approximate the true DGP with a model \(\hat{p}(y|x)\) using training data when training an SML model. The approximation introduces estimation error
\begin{equation}
\epsilon_i = \hat{\boldsymbol \pi}_i - \boldsymbol \pi_i = \hat{p}(y_i|x_i) - p(y_i|x_i^{\text{true}})\,.
\label{eq_est_error}    
\end{equation}

Algorithmic fairness literature often assumes the absence of estimation error~\cite[see e.g.,][]{hardt2016} or assumes that the true causal structure is known~\cite{Zhang2017,kilbertus2018,chiappa2019,carey2022causal}.
In practice, this is rarely the case. 
Hence, it is crucial, both practically and legally, to distinguish between the true underlying probabilities \(\boldsymbol \pi_i\) and the estimated probabilities \(\hat{\boldsymbol \pi}_i\).
While disparities in the true underlying probability may sometimes be legitimate or justified (Section~\ref{sec_true_data}), introducing an estimation error that disadvantages individuals based on protected attributes invokes discrimination liability. 

As the model will try to approximate the true DGP, modellers' expectations are difficult to ascertain. 
Law is unlikely to set a deterministic standard that any adverse effects of estimation will make a modeller liable, yet the modeller should try to approximate the true model as well as possible~\cite[see][for discussions on model misspecification]{akaike1973information,white1982maximum,vehtari2012survey,vehtari2017practical}. 
Where an estimation disparity reaches a threshold for discriminatory effects, the legal evaluation would require analysing the steps taken to test and mitigate estimation disparity (even though the intent is immaterial). 

The potential bias in training data presents a risk that the estimation model will introduce bias against individuals with protected attributes (Section~\ref{sec_legitimate_y}).
Historical discriminatory lending practices, for example, could be perpetuated through biased training data~\cite{black1978,rice2013,sargeant2023,sargeant2025}. 
Such biased estimations may introduce biased outcomes that are not reflective of true differences, potentially leading to discriminatory outcomes.
Therefore, we introduce ``Conditional Estimation Parity'' to formalise the legal context of estimation.

\textbf{Conditional Estimation Parity} is the difference in estimation error between groups with a protected attribute, given legitimate features, i.e.,
\begin{equation}
\mathbb{E}_x[\epsilon \mid x_p, x_l] = \mathbb{E}_x[\epsilon \mid x_l]\,.
\end{equation}
Reducing the error in Eq.~\ref{eq_est_error} is expected to diminish the risk of conditional estimation disparity. 
However, assessing conditional estimation parity is complex due to inherent challenges in evaluating estimation error. 
It is crucial to examine both mathematical and legal causal theories of why certain differences are legitimate bases to make classification distinctions~\cite{kiviat2023}.
Assuming a model action appears to be prohibited by the rule articulation function, the legal analysis shifts to whether liability can be proven. 
We now examine the basis for identifying statistical disparities and legitimate differentiation in the context of unlawful discrimination. 

\section{How can a decision-maker be liable for unlawful discrimination?}\label{sec:liable}

\subsection{Statistical Disparities and \textit{Prima Facie} Discrimination} \label{sec_discriminatory_effect}

To initiate a claim for direct or indirect discrimination, a claimant must establish a \textit{prima facie} case~\cite[s 136]{EWCA2005,EqualityAct}.

In direct discrimination, the claimant must show the alleged discrimination explicitly referred to a protected characteristic or exact proxy; the protected characteristic has to be the reason for the less favourable treatment~\cite{court2018,court2017}.
For indirect discrimination, sufficient evidence must be produced to identify the PCP, identify the protected group, and show that the PCP places the protected group at a particular disadvantage when compared to those without such attribute~\cite{EWCA2005,Essop}.
Although courts do not always defer to statistical evidence~\cite{court2012}, it is commonly used and will likely be essential in an algorithmic context.

Some algorithmic fairness papers have written that the US law defines explicit statistical thresholds to define a \textit{prima facie} case, such as the four-fifths rule in employment discrimination law~\cite{romei2014,feldman2015,zafar2017}.
However, this conflates the legal position of these tests, which are certainly not rigid definitions of differential outcomes~\cite{EEOC,Watkins2024,Harned2020}.
In any event, UK law is more flexible than the US on thresholds for statistical significance which are often resisted by courts to avoid excessive dependence on data~\cite{Seymour,sanchez2020}.
Statistical significance depends on the context of the comparison~\cite{Essop,Seymour}, and smaller disparities are less likely to trigger legal inquiry under anti-discrimination laws~\cite{Essop}.

Statistical disparities may indicate a reason to consider whether discrimination has arisen. 
However, without taking context and potential true and legitimate differences into account, these disparities hold little legal weight (Section \ref{sec:legitimacy_of_true_differences}).
We can formalise this as the legal target being to minimise the conditional estimation disparity
\begin{equation}
\omega = |\mathbb{E}_x\left[\epsilon \mid x_l, x_p\right] - \mathbb{E}_x\left[\epsilon\mid x_l\right]\,|, 
\label{eq_minimize_stat_disparity}
\end{equation}
where $|\cdot|$ denotes the absolute value. 
This target generalises the idea of minimising conditional statistical parity. 
If we assume \emph{true} conditional statistical parity, i.e. 
\begin{equation}
\mathbb{E}_x\left[p(y|x^{\text{true}}) \mid x_l, x_p\right] = \mathbb{E}_x\left[p(y|x^{\text{true}})\mid x_l\right]\,, 
\label{eq_true_stat_disparity}
\end{equation}
then the target in Eq.~\ref{eq_minimize_stat_disparity} will be reduced to minimise the conditional statistical parity (see Eq.~\ref{eq_cond_stat_parity}). 

Hence, if true statistical parity does not hold, it is explained by true differences between groups. 
If there is a true difference, such as age in financial services, forcing conditional statistical parity could harm the protected group, most likely resulting in unlawful discrimination. 
This result aligns with previous observations about the risks of forcing parity metrics \cite{corbett2017,xiang2021,hertweck2021,sargeant2025}. 
Courts may need to be more flexible in the type of statistical data they consider to establish a \textit{prima facie} case by considering non-comparative adverse effects in their assessment.
Therefore, deferring to conditional estimation parity provides an avenue for a contextually informed~assessment. 

\subsection{Legal Causation for Unlawful Discrimination}
\label{sec_causation_utility}

To lawyers, causation is the relationship between an act and its effect, which requires two questions: (1) factually, \textit{but for} the act, would the consequences have occurred; (2) is the act a substantial cause of the consequence to apply legal responsibility.
We are concerned with the first question.
Direct discrimination ``requires a causal link between the less favourable treatment and the protected characteristic''; indirect discrimination ``requires a causal link between the PCP and the particular disadvantage suffered by the group and individual [sharing the protected characteristic]''~\cite{Essop}.

There is ongoing debate about whether legal causation poses an insurmountable obstacle in proving algorithmic discrimination, given the technical difficulties in establishing direct causal links between a protected characteristic and the resulting outcome.
Some scholars highlight distinct and higher standard of causation under US law~\cite[discussing the heightened ``robust causality requirement'' introduced in US Supreme Court case, \textit{Inclusive Communities}]{hurley2017,barocas2016}, while some focus narrowly on statistical causality and correlations, which do not always align with legal causation~\cite{xenidis2020,williams2018}.
While it is true even in the UK that ``a correlation is not the same as a causal link''~\cite{Essop}, courts have shown a functional flexibility in the application of causation in order to navigate complex human decision-making processes to determine legal causation~\cite{EandJFS}, which can be even more intricate or opaque than algorithmic processes. 

In an algorithmic context, this causal link requires asking whether \(i\) would have received the same action or decision \(a\), \textit{but for} their protected attribute \(x_p\) or the PCP that indirectly relates to their protected attribute \(x_p\)~\cite{lords1989,lords1990}.
From a decision-theoretic perspective, the protected attribute \(x_p\) can affect the decision \(a\) either through the utility function \(u(a,y)\) or through the model \(\hat{p}(y|x)\). 
Discrimination may occur if the utility function in Eq.~\ref{eq_optimal_action} differs for different groups defined by the protected attribute. 
This concept parallels taste-based discrimination, where a decision-maker's subjective preferences or prejudices against a group lead to differences in outcomes~\cite{becker1957}.
In this context, the utility function \(u(a,y)\) unjustifiably disfavours a group based on protected attributes \(x_p\). 
Such a difference would mean that an individual or whole group with a protected attribute is treated less favourably than those without a protected attribute given the same model \(\hat{p}(y|x)\).
Such a difference in the utility function would risk unlawful discrimination. 
Specifically, if \(u(a,y)\) is changed for different persons, either \textit{directly} based on a protected attribute or it \textit{indirectly} has the effect of disproportionately disadvantaging a group with a protected characteristic without justification.
A detailed legal assessment would consider the specifics of the case to determine legal causation.

Differing \(\hat{p}(y|x)\), on the other hand, would mean that there is an indirect causation between the decision \(a\) and \(x_p\). 
The differing \(\hat{p}(x|y)\) is analogous to statistical discrimination, where group-level statistics are used as proxies for individual characteristics due to imperfect information~\cite{arrow1971,phelps1972}.
This might either be motivated by true differences or a result of conditional estimation disparity. 
In the latter case, this might be a case of legal causation, i.e., that the model is poor, and hence, the modelling has resulted in disadvantaging a protected group.
Therefore, we can view the legal causal structure of \(\hat{p}(y|x)\) as central to avoiding unlawful discrimination. 
However, not considering legal causation could lead to conditional estimation disparity, and potentially result in unlawful discrimination.

Legal causation focuses on the causal link between \(x_p\) and the decision \(a\). 
Additionally, legal causation is less formal than common definitions of causal effects in ML. 
Courts, at least outside of the US, are effects-orientated, and a wide range of forms of a ``legal causal link'' could be identified~\cite{shin2013,khaitan2015}. 
Much of the causal-based fairness literature formulates ``causation'' on the true causal model structure in \(\hat{p}(y|x)\), i.e., the study of the causal effect of \(x\), due to outside interventions on \(y\)~\cite{Pearl2010,barocas2016,zhang2018fairness,carey2022causal}. 
However, this formulation is not the same as that of legal causation.

\section{When can a decision-maker be excused for unlawful discrimination?} \label{sec:defences}

\subsection{No Defence for Direct Discrimination}
A unique feature of UK direct discrimination is that: ``In contrast to the law in many countries, where English law forbids direct discrimination it provides no defence of justification''~\cite{EandJFS}.
In the UKSC, Lord Phillips explained that while ``it is possible to envisage circumstances where giving preference to a minority racial [or other protected] group will be justified'' nevertheless ``a policy which directly favours one racial [or other protected] group will be held to constitute racial discrimination against all who are not members of that group''~\cite{EandJFS}.
Lady Hale confirmed that ``however justifiable it might have been, however benign the motives of the people involved, the law admits of no defence''~\cite{EandJFS}.
Given direct algorithmic discrimination is more likely to arise in the UK, the absence of a defence is significant. 
It is important to correct misconceptions that direct discrimination are either unlikely to arise or may be excused, such as in the US where a legitimate, non-discriminatory justification can excuse disparate~treatment.

\subsection{Potential Justification of Indirect Discrimination} 
\label{sec_legitimate_y}

On the other hand, indirect discrimination can be excused if the PCP is objectively justified as a proportionate means of achieving a legitimate aim~\cite[s 19(2)(d)]{EqualityAct}.
Decision-makers must consider the legitimacy of using an SML model by explicitly defining its purpose and the outcome variable \(y\). 
In algorithm design, social implications should be considered~\cite{mulligan2019,hu2020,kasy2021}, as well as alignment with legal~expectations.

\subsubsection{Defining legitimate aims \(y\)}
Identifying a legitimate aim is closely connected to the choice of \(y\), the unknown entity used for decision-making. 
The legitimacy of the aim depends on the decision-makers' \textit{raison d'être}~\cite{khaitan2015,justice1986,court2006}. 
In \textit{Homer}, the UKSC established a legitimate aim must ``correspond to a real need and the means used must be appropriate with a view to achieving the objective and be necessary to that end''~\cite{court2012}. 
For example, in lending, it is a legitimate aim to protect the repayment of their loans or at least secure their loans. 
In fact, ``the mortgage market could not survive without that aim being realised''~\cite{bristol2015}.  
If the choice of \(y\) is legitimate based on context and the benefit outweighs any potential harm, there is a lower risk of unlawful discrimination~\cite{justice1986}.

For a legitimate~\(y\)~to be an exception to indirect discrimination, the PCP must be a proportionate means of achieving the legitimate~\(y\)~\cite{EqualityAct}. 
To be proportionate, it must be an appropriate means of achieving the legitimate aim and (reasonably) necessary to do so~\cite{court2012}. 
Such analysis will turn on the facts of each case by evaluating whether the design choices were ``appropriate with a view to achieving the objective and be necessary'' and weighing the need against the seriousness of detriment to the disadvantaged group~\cite{court2006}. 
Proportionality assessments balance the discriminatory effect of a PCP with the reasonable needs of a business~\cite{lane2018}, and consider whether non-discriminatory alternatives were available~\cite{court2012}. 
While important work on identifying less discriminatory models is emerging~\cite{finreglab_machine_2023,gillis2024,black2024}, such efforts should be viewed within the broader context. 
A less discriminatory \textit{algorithm} should not substitute pursuing less discriminatory \textit{alternatives}, which may involve dispensing with algorithmic decision-making altogether. 
Measures to improve accuracy, maximise benefits over costs, minimise estimation error, or condition for protected attributes may all be relevant considerations for whether the modeller's choices were proportionate means of achieving a legitimate~\(y\).

It is not always possible to use \(y\) directly, instead we use an approximation \(\tilde{y}\) of the true underlying \(y\), which can lead to biased predictions. Let
\begin{equation}
\gamma_i = ||p(\tilde{y}_i|x_i^{\text{true}}) - p(y_i|x_i^{\text{true}})||_2\,,
\label{eq_est_bias}    
\end{equation}
then, if the expectation of \(\gamma\) condition on \(x_l\) shows a disparity, i.e.,
\begin{equation}
    \mathbb{E}_x[\gamma \mid x_p, x_l] \neq \mathbb{E}_x[\gamma \mid x_l]\,,
\end{equation}
it suggests the use of \(\tilde{y}\) is inappropriate and might be discriminatory.

The approximation of the true target variable \(\tilde{y}\) can introduce target-construct mismatch, meaning the proxy target variable does not fully align with the underlying phenomenon it aims to measure~\cite{jacobs2021,wang2024}.
To illustrate with an example, if a bank's training data is outdated or sourced from a different country, it may not accurately represent the current population relevant to the model. 
This discrepancy can lead to biased estimates, particularly if the data reflects historical prejudices. 
For instance, the model might unjustly associate certain demographics with higher default risk, not because of true differences in \(y\) but biased data in \(\tilde{y}\)~\cite[as warned in][]{dfs2021}.

\subsubsection{Defining legitimate and non-legitimate variables \(x\)}\label{sec_legitimate_x}

Modellers need to examine individual features to ensure its use is legitimate or not. 
Defining legitimate variables \(x_l\) and non-legitimate variables \(x_n\) similarly draws on context and relationship with the true DGP. 
We define a non-legitimate feature is one that, if included, would not be included in the true DGP and hence would lack legal causal effects.
%contribute to the predictive performance of the optimal model, i.e., the one with the lowest estimation error (Section~\ref{sec_estimation_parity}).
Therefore, \(x_n\) would not improve the predictive performance if a modeller had the true features.

Traditional guidance often advocates incorporating all available data to maximise predictive accuracy, typically without explicit consideration of causal links between model inputs and the predicted outcomes~\cite{berk2019,manski2022,manski2023}. 
However, in the presence of label bias, measurement error, or other issues arising from flawed data-generating processes, these approaches risk introducing or amplifying algorithmic bias~\cite{obermeyer2019,jacobs2021,zanger2024,mikhaeil2024}. 
In a medical context, \citet{obermeyer2019} highlight the importance of modifying the data labels provided to algorithms, emphasising that this process demands deep domain-specific knowledge, iterative experimentation, and careful selection of labels that genuinely represent the true outcomes of interest.
For legally compliant models, it is crucial to move beyond naïve performance maximisation toward more nuanced, context-sensitive approaches. 
We argue that deliberate feature selection should explicitly reflect the causal relationships between model inputs, the target variable, and the underlying true DGP. 

For example, in lending, hair length strongly correlates to gender in many cultural contexts but is unlikely to contribute to a consumer's true default risk. 
\citet{boyarskaya2022} explain the absence of a ``causal story'' between hair length and loan repayment because hair length would not be part of a true model for the risk of default. 
Therefore, hair length is an example of~\(x_n\)~in a lending context. 

For comparison, the legitimacy of zip codes illustrates the nuanced nature of legitimate features. 
While a zip code may correlate with race in some contexts, it might be a legitimate variable in other situations. 
For example, in an application for home insurance covering flood risk, zip codes are invaluable proxies for granular information such as geographical features, land topography, and historical flooding. 
Therefore, in the best model for property flood insurance decisions, zip code will improve the predictive performance as a legitimate proxy for the true geographical features within the true DGP.
However, in a university application, zip code should not be predictive or causal to a prospective student's merit for acceptance.
In such cases, zip code likely acts as a proxy for race or the unprotected characteristic of socio-economic status and would be \(x_n\).
So, in some circumstances, the zip code would be legitimate \(x_l\), but in others, it may not be legitimate \(x_n\).
It will also be relevant to consider whether a less discriminatory feature is available, i.e., one with less correlation to a protected attribute that is equally predictive.

As explored in Section~\ref{sec_case_study}, in lending predictions information about income, employment, and debts are likely to be legitimate features~\(x_l\).
Credit scores, or related features, would have a material impact on the true model for default, and then would be a legitimate feature~\(x_l\)~\cite{black1978,hurley2017}.
Given that nearly all features may contain some information on protected attributes, even legitimate factors~\cite{corbett2023}, this approach explains the need to assess the strength of this dependence and whether the feature contributes significantly to the model's prediction and can be argued to be part of a true DGP. 

\section{Case Study} \label{sec_case_study}

To demonstrate a real-world approach to the legal reasoning through our formalisation, we now discuss our framework by reference to the first case regarding automated decision-making and discrimination, decided by the National Non-Discrimination and Equality Tribunal of Finland (\textbf{Tribunal})~\cite{finland}.
At the time of writing, however, there are no reported UK court decisions that scrutinise algorithmic discrimination in an equivalent setting, and only a handful of such judgments exist worldwide (even fewer outside the United States).
Given Finnish anti-discrimination law bears many similarities to UK and EU laws, we use this comparative case analysis to concretely illustrate our framework rather than hypothesise.
In Appendix~\ref{finnishlaw}, we include the relevant provisions of the Finnish Non-Discrimination Act to demonstrate the similarities to the UK Equality Act.

In this case, Person A, was denied credit for online purchases based on a credit rating system employed by a bank. 
The Tribunal found that the bank's statistical scoring model resulted in discrimination based on multiple protected characteristics and was not justified by an acceptable objective achieved by proportionate measures. 
Consequently, the Tribunal prohibited the bank from continuing this practice and imposed a conditional fine to enforce compliance.

\subsection{Decision-Making Model and Data}

The decision-making system in question is for online store financing credit. 
The credit applied for by the consumer in each situation is also always bound to the purchase and its value, which means that it is more difficult, or even impossible, to undertake detailed requests for information and background checks. 
The credit decisions were based on data from the credit company's internal records, credit file information, and the score from the company's internal scoring system.
The bank's scoring system assessed creditworthiness.
The scoring system used population statistics and personal attributes to calculate the percentage of people in certain groups with bad credit history and awarded points proportionate to how common bad credit records were in the group in question.
The variables used included race, first language, age, and place of residence. 
The scoring system did not require or investigate the applicant's income or financial situation. 

\subsection{True Data Generating Process and Estimation Error}

The bank's scoring model was based on multiple variables where the majority were protected attributes, including gender, language, age and place of residence, meaning the model is more or less~\(\hat{p}(y|x_p)\).

This model did not attempt to model the true underlying DGP and instead relied on data that was available. 
It is reasonable to expect that the bank was aware of other legitimate factors that could explain the credit score. 
Therefore, the model lacks information that could have been used to make better predictions, i.e., legitimate features~\(x_l\).
By solely using the data available, rather than identifying data that would be best to reduce conditional estimation error, the modellers built an automated decision-making system that unlawfully discriminated. 
We now evaluate how the Tribunal came to those conclusions about the legitimacy of~\(y\)~and~\(x\)~for~such~a~model. 

\subsection{Legitimate \(y\)}

The bank argued that the ``different treatment does not constitute discrimination if the treatment is based on legislation and has an otherwise acceptable objective and the measures to attain the objective are proportionate.''
The Tribunal agreed that ``the provision of credit to customers is a business,
the purpose of which is to gain profit'' and that ``the investigation of creditworthiness is as such based on law and that it has the acceptable and justified objective as defined in section 11 of the Non-Discrimination Act''.
Therefore, creditworthiness assessment is a legitimate \(y\) in this context. 

However, the Tribunal clarified that the creditworthiness assessment ``means expressly the assessment of an individual’s credit behaviour, credit history, income level and assets, and not the extension of the impact of models formed on the basis of probability assessments created with statistical methods using the behaviour and characteristics of others, to the individual applying for the credit in the credit decision in such a way that assessment is solely based on such models.''
To be appropriate and necessary to achieve that aim, therefore, the model must use legitimate features~\(x_l\).

\subsection{Legitimate, Non-Legitimate, and Protected Variables \(x\)}

\noindent
The Tribunal evaluated each input variable against two cumulative tests:  
(i)~its factual relevance to an individual’s likelihood of repayment and  
(ii)~its permissibility under fundamental-rights and sector-specific law.  
Variables that clear both hurdles can be labelled \emph{legitimate} (\(x_l\)); those that fail the predictive-relevance test are \emph{non-legitimate} (\(x_n\)); and those barred by anti-discrimination law, even if potentially predictive, form the set of \emph{protected variables} (\(x_p\)).

\paragraph{Legitimate Variables \(x_l\)}

As explained by the Tribunal, to achieve the legitimate \(y\) of undertaking an individual assessment of creditworthiness, the model should have considered, for example, income, expenditure, debt, assets, security and guarantee liabilities, employment and type of employment contract (i.e., permanent or temporary). 
These features would have been legitimate variables~\(x_l\)~by improving the predictive performance of the model for the individual.

\paragraph{Non-Legitimate Variables \(x_n\)}

Four protected characteristics~\(x_p\)~were used as variables in this model. 
The Tribunal considered whether the use of these protected attributes was legitimate. 
Unlike the economic variables listed above, the following attributes either lacked a demonstrable causal link to repayment or were expressly prohibited~by law.

\textbf{Age} was a non-legitimate variable in this context, but the Tribunal acknowledged that age may be a legitimate variable if it had been used in the assessment of creditworthiness of young persons with limited credit history. 
Age did not contribute to model accuracy in a way that could be argued as part of the true DGP.  

\textbf{First language} was a non-legitimate variable in the credit assessment because it ``will result, de facto, in the segregation on ethnic lines, the justification of which does not include compelling arguments that could be deemed acceptable from the point of view of the system of fundamental rights.'' 
Evidence showed the model ranked Finnish-speaking residents lower than Swedish-speaking residents. 
Further, ethnic minorities with another official first language were put in a more unfavourable position. 

\textbf{Place of residence} was a non-legitimate variable because the bank had not provided any empirical evidence that an ``assumption made on the basis of the general data of residences in a certain area would prove anything of the loan repayment capacity of an individual resident in that area.'' Absent such evidence, postcode information remains no more than a group-level stereotype and therefore fails both Tribunal tests.

\textbf{Gender} could not be a legitimate variable because it is prohibited from being used as an actuarial factor in financial services under EU law~\cite{TestAchats}, also implemented in UK law~\cite{EA2012}.
The express legal prohibition, in this context, means the Tribunal considered that gender should not be part of the true DGP. 
Although it was shown that women received a higher score than men and that if Person A had been a woman, he would have been granted the credit. 
Therefore, the Tribunal's approach demonstrates that statistical disparities alone cannot override normative and legal prohibitions.

Therefore, in this case, some of the protected variables \(x_p\) cannot be used in the model, whereas the protected characteristic of age may sometimes be considered a legitimate variable \(x_l\).

\subsection{Conditional Estimation Parity}

Building upon the legitimate variables identified above, we now examine the concept of conditional estimation parity, which pertains specifically to disparity in estimation error between groups distinguished by a protected attribute, conditional on legitimate features. 
Reducing the estimation error articulated in Eq.~\ref{eq_est_error} thus mitigates the risk of conditional estimation disparity.
However, directly assessing conditional estimation parity poses challenges due to the inherent difficulties in measuring estimation error.

In our case study, the Ombudsman presented evidence of the effects of the protected characteristics \(x_p\) on the true prediction. 
For example, evidence showed that even when conditioned on legitimate variables, the model continued to rank Finnish-speaking and Swedish-speaking residents differently.
This observation demonstrates a violation of conditional statistical parity (Eq.~\ref{eq_cond_stat_parity}), as the predictions differ for groups distinguished by protected attributes after conditioning on legitimate variables.
Additionally, because the Tribunal concluded there should be no legitimate differences in creditworthiness between these protected groups, the observed difference also indicates a violation of conditional estimation parity (Eq.~\ref{eq_est_error}). 
The difference arises specifically from the estimation error -- the discrepancy between the true underlying parity (as established by the Tribunal) and the predictions generated by the model.
The relevant legal evaluation is then whether it is a valid true difference or, as in this case, it is based on a protected characteristic~\(x_p\).
Therefore, in this scenario, violations of conditional statistical parity and conditional estimation parity are closely related -- the presence of estimation error directly leads to differential outcomes in model predictions. 
This highlights the importance of understanding the origins of prediction disparities through the lens of estimation errors, which can guide legal and practical evaluation of fairness and responsibility in predictive models.

Judicial legal reasoning is complex and requires engaging with this type of reasoning through statistical or theoretical means.
Unlike other work that proposes quantitive thresholds or metrics for legality, we focus on formalising modelling choices within UK anti-discrimination law. 
This approach allows us to outline a flexible, decision-theoretic framework, aligning modeller and decision-maker choices as optimisation under legal constraints.
Such a structured yet legally-informed framework aligns more closely with judicial reasoning, representing a fundamental and novel contribution to the literature on algorithmic discrimination.

\section{Conclusion} \label{sec:conclusion}

This paper contributes to the understanding of unlawful discrimination in SML under UK law and related jurisdictions such as the EU and across the Commonwealth. 
Unlike the statistical focus in much of the algorithmic fairness literature, we present a legally grounded, decision-theoretic framework centred on the true data-generating process and its connection to legal causation. 

By emphasising the legitimacy of prediction targets and features, we demonstrate how legally informed concepts, like conditional estimation parity, offer a more cohesive means of assessing when and how models discriminate. 
By situating these considerations within a UK context that shares key principles with many other jurisdictions, our approach enables more robust global approach to legal doctrine. 

Crucially, our work corrects several mischaracterisations around discrimination, including those stemming from literature that misrepresents US law, fostering assumptions in the field that are not legally-informed, and literature that accurately reflects US law but diverges materially from UK law.

\paragraph{Recommendations}
Minimising unlawful discrimination in automated decision-making requires a nuanced and contextual approach. 
Our findings underscore several key considerations to identify and mitigate potential discrimination:
\begin{enumerate} 
\item \emph{Assess data legitimacy}. Carefully examine if the data, both the target variable (\(y\)) and features (\(x\)), are legitimate for the specific context (Sections~\ref{sec_legitimate_y} and \ref{sec_legitimate_x}). Legal analysis should inform what is legitimate in a specific setting.
\item \emph{Build an accurate model}. Aim to approximate the true DGP \(p(y|x)\), using only legitimate features~\(x_l\). Take reasonable, necessary, and proportionate steps to minimise estimation error and aim for estimation parity~(Section~\ref{sec_estimation_parity}). 
\item \emph{Evaluate differences}. Given the best model \(\hat{p}(y|x)\), assess for conditional statistical parity by examining outcomes across groups with protected characteristics (Section~\ref{sec_discriminatory_effect}). If differences persist, consider whether these disparities are legitimate variations. Practitioners should incorporate further legitimate features to mitigate these disparities or, if necessary, refrain from deploying the model to prevent unlawful discrimination.
\end{enumerate}

These steps, informed by our legal and statistical framework, offer a structured approach to designing, training, and auditing SML models to reduce the risk of unlawful discrimination before it arises in automated decisions.

\paragraph{Limitations}

We acknowledge the following two limitations. 
First, our paper is formally limited to analysing and providing novel recommendations for the UK and related jurisdictions. 
While comparative research is valuable, the minimal UK-specific research on unlawful algorithmic discrimination necessitates a focused approach. 
We discuss related jurisdictions that are functionally similar and based on English common law, and draw comparisons from different jurisdictions where appropriate.
However, our paper emphasises the necessity of careful classification by experts with appropriate and jurisdictional-specific legal advice.
Second, our paper's contributions are theoretical from a legal perspective. 
Although we recognise the importance of applied work, the primary aim is to establish a new theoretical framework that does not exist in the literature. 
To bridge the gap between theory and application, our case study illustrates how the framework could be implemented, offering insights into its potential real-world utility.
%While this work provides a conceptual foundation, it is not a substitute for legal expertise or actionable tools but serves to guide researchers and practitioners in aligning ML models with legal fairness standards. Instead, it provides a conceptual foundation and methodological principles that can guide both researchers and practitioners, ensuring that algorithmic interventions in high-stakes decision-making meet not just technical metrics of fairness, but also stand on firm legal ground.

In conclusion, this work bridges a critical gap between the technical aspects of automated decisions and the complexities of anti-discrimination law. By translating these nuanced legal concepts into decision theory, we underscore the importance of accurately modelling true data-generating processes and the innovative concept of estimation parity. Our approach enhances the understanding of automated decision-making and sets a foundation for future research that aligns technological advancements with jurisdictional-specific legal and ethical standards.

%\begin{acks}
%\end{acks}

\newpage
\bibliographystyle{ACM-Reference-Format}
\bibliography{references}

\appendix

\section{Overview of Finnish Anti-Discrimination Law} \label{finnishlaw}

Person A reported their case to the Non-Discrimination Ombudsman (Yhdenvertaisuusvaltuutettu), who brought the case before the National Non-Discrimination and Equality Tribunal (Yhdenvertaisuus-ja tasa-arvolautakunta).
The Finnish Non-Discrimination Act is the relevant law~\cite{FinlandNDA}. 
Important extracts are quoted here using the official English translation, although only the Finnish and Swedish (not included) language is legally binding.

Section 8(1) of the Non-Discrimination Act defines the protected characteristics as:
\begin{quote}
    \textit{No one may be discriminated against on the basis of age, origin, nationality, language, religion, belief, opinion, political activity, trade union activity, family relationships, state of health, disability, sexual orientation or other personal characteristics. Discrimination is prohibited, regardless of whether it is based on a fact or assumption concerning the person him/herself or another.} 
\end{quote}

\begin{quote}
    \textit{Syrjinnän kielto Ketään ei saa syrjiä iän, alkuperän, kansalaisuuden, kielen, uskonnon, vakaumuksen, mielipiteen, poliittisen toiminnan, ammattiyhdistystoiminnan, perhesuhteiden, terveydentilan, vammaisuuden, seksuaalisen suuntautumisen tai muun henkilöön liittyvän syyn perusteella. Syrjintä on kielletty riippumatta siitä, perustuuko se henkilöä itseään vai jotakuta toista koskevaan tosiseikkaan tai oletukseen.}
\end{quote}

Section 3(1) of the Non-Discrimination Act provides that: ``Provisions on prohibition of discrimination based on gender and the promotion of gender equality are laid down in the Act on Equality between Women and Men (609/1986).''
The Non-Discrimination Act can be applied in cases of multiple discrimination, even if gender is one of the grounds of discrimination~\cite[][s 3(1)]{FinlandGender,FinlandNDA}.

It is worth noting that this definition is broader than in the UK Equality Act. Some protected characteristics are outlined more explicitly; for example, a person discriminated against on the basis of language may be able to bring a claim based on racial discrimination~\cite{dziedziak2012}. 
Unlike many Nordic countries, the UK Equality Act does not explicitly protect political activity, trade union activity, and does not include ``or other personal characteristics''~\cite{hellum2023}.

Direct discrimination is defined in Section 10:
\begin{quote}
    \textit{Discrimination is direct if a person, on the grounds of personal characteristics, is treated less favourably than another person was treated, is treated or would be treated in a comparable situation.}
\end{quote}

\begin{quote}
    \textit{Syrjintä on välitöntä, jos jotakuta kohdellaan henkilöön liittyvän syyn perusteella epäsuotuisammin kuin jotakuta muuta on kohdeltu, kohdellaan tai kohdeltaisiin vertailukelpoisessa tilanteessa.}
\end{quote}

Indirect discrimination is defined in Section 13:
\begin{quote}
    \textit{Discrimination is indirect if an apparently neutral rule, criterion or practice puts a person at a disadvantage compared with others as on the grounds of personal characteristics, unless the rule, criterion or practice has a legitimate aim and the means for achieving the aim are appropriate and necessary.}
\end{quote}

\begin{quote}
    \textit{Syrjintä on välillistä, jos näennäisesti yhdenvertainen sääntö, peruste tai käytäntö saattaa jonkun muita epäedullisempaan asemaan henkilöön liittyvän syyn perusteella, paitsi jos säännöllä, perusteella tai käytännöllä on hyväksyttävä tavoite ja tavoitteen saavuttamiseksi käytetyt keinot ovat asianmukaisia ja tarpeellisia.}
\end{quote}

Section 11(1) defines justifications for different treatment as:
\begin{quote}
    \textit{Different treatment does not constitute discrimination if the treatment is based on legislation and it otherwise has an acceptable objective and the measures to attain the objective are proportionate.}
\end{quote}

\begin{quote}
        \textit{Erilainen kohtelu ei ole syrjintää, jos kohtelu perustuu lakiin ja sillä muutoin on hyväksyttävä tavoite ja keinot tavoitteen saavuttamiseksi ovat oikeasuhtaisia.}
\end{quote}

\end{document}